\documentclass[12pt]{article}
\usepackage[colorlinks]{hyperref}
\usepackage{xcolor}
\usepackage[T1]{fontenc}
\usepackage{amsmath,amsfonts,amsthm,amssymb}
\usepackage{relsize}
\usepackage{bm}
\usepackage{epsfig}  
\usepackage{subcaption}
\def\ftoday{{\sl {Le \number\day \space\ifcase\month 
\or janvier\or f\'evrier\or mars\or avril\or mai
\or juin\or juillet\or ao\^ut\or septembre\or octobre
\or novembre \or d\'ecembre\fi\space \number\year}}}    
\def\ptoday{{\sl {\number\day \space de\space \ifcase\month 
\or janeiro\or fevereiro\or mar{\c c}o\or abril\or maio
\or junho\or julho\or agosto\or setembro\or outubro
\or novembro \or dezembro\fi\space de\space \number\year}}}    
\def\gtoday{{\sl {Den \number\day. \ifcase\month 
\or Januar\or Februar\or M\"arz\or April\or Mai
\or Juni\or Juli\or August\or September\or Oktober
\or November \or Dezember\fi\space \number\year}}}    
\def\today{{\sl {\ifcase\month
\or January\or February\or March\or April\or May
\or June\or July\or August\or September\or October
\or November \or December\fi \space\number\day,\space 
                                            \number\year}}}
\newcommand{\journal}[4]{{\em #1~}#2\,(#3)\,#4}

\newcommand{\ijmp}{\journal {Int. J. Mod. Phys.}}

\newcommand{\pr}{\journal {Phys. Rev.}}

\newcommand{\sovietjetpl}{\journal {Sov.Phys.JETP}}
\newcommand{\prl}{\journal {Phys. Rev. Lett.}}

\newcommand{\rmp}{\journal {Rev. Mod. Phys.}}
\newcommand{\cmp}{\journal {Commun. Math. Phys.}}

\newcommand{\np}{\journal {Nucl. Phys.}}

\newcommand{\prep}{\journal {Phys. Rep.}}

\newcommand{\nc}{\journal {Nuovo Cimento}}

\newcommand{\annp}{\journal {Ann. Phys. (N.Y.)}}

\renewcommand{\a}{\alpha}
\renewcommand{\b}{\beta}
\newcommand{\g}{\gamma}           
\renewcommand{\d}{\delta}         \newcommand{\D}{\Delta}
\newcommand{\e}{\epsilon}
\newcommand{\ka}{\kappa}
\newcommand{\la}{\lambda}        
\newcommand{\m}{\mu}
\newcommand{\n}{\nu}
\newcommand{\om}{\omega}         
              
\newcommand{\s}{\sigma}           
\newcommand{\f}{{\phi}}           

\renewcommand{\AA}{{\cal A}}
\newcommand{\BB}{{\cal B}}
\newcommand{\CC}{{\cal C}}

\newcommand{\FF}{{\cal F}}
\newcommand{\GG}{{\cal G}}

\newcommand{\NN}{{\cal N}}

\newcommand{\TT}{{\cal T}}

\newcommand{\XX}{{\cal X}}

\newcommand{\es}{\\[3mm]}

\newcommand{\sla}{\raise.15ex\hbox{$/$}\kern -.57em} 
\newcommand{\Sla}{\raise.15ex\hbox{$/$}\kern -.70em}

\def\Lac{\displaystyle{\Bigl\{}}
\def\Rac{\displaystyle{\Bigr\}}}
\newcommand{\lp}{\left(}\newcommand{\rp}{\right)}

\newcommand{\complex}{{\kern .1em {\raise .47ex
\hbox {$\scriptscriptstyle |$}}
    \kern -.4em {\rm C}}}
\newcommand{\real}{{{\rm I} \kern -.19em {\rm R}}}
\newcommand{\rational}{{\kern .1em {\raise .47ex
\hbox{$\scripscriptstyle |$}}
    \kern -.35em {\rm Q}}}
\renewcommand{\natural}{{\vrule height 1.6ex width
.05em depth 0ex \kern -.35em {\rm N}}}

\newcommand{\tr}{{\rm {Tr} \,}}
\newcommand{\half}{\dfrac{1}{2}}
\newcommand{\pa}{\partial}

\newcommand{\eg}{{\em e.g.,\ }}

\newcommand{\ie}{{{\em i.e.},\ }}
\newcommand{\Ie}{{\em I.e.,\ }}

\newcommand{\etc}{{\em etc.\ }}

\newcommand{\twiddle}{\lower.9ex\rlap{$\kern -.1em\scriptstyle\sim$}}

\newcommand{\equ}[1]{(\ref{#1})}
\newcommand{\eq}{\begin{equation}}
\newcommand{\eqn}[1]{\label{#1}\end{equation}}
\newcommand{\eea}{\end{eqnarray}}
\newcommand{\eqa}{\begin{eqnarray}}
\newcommand{\eqan}[1]{\label{#1}\end{eqnarray}}
\newcommand{\ba}{\begin{array}}
\newcommand{\ea}{\end{array}}
\newcommand{\eqac}{\begin{equation}\begin{array}{rcl}}
\newcommand{\eqacn}[1]{\end{array}\label{#1}\end{equation}}

\makeatletter 
\@addtoreset{equation}{section}  
\makeatother 
 
\newcommand{\ben}{\begin{enumerate}}
\newcommand{\een}{\end{enumerate}}

\newcommand{\cyclsum}{\mathlarger{\mathlarger{\circlearrowleft}}}

\newcommand{\scc}{s_{(c)}}
\newcommand{\Tup[1]}{\overset{(#1)}{T}}

\newcommand{\bi}{{\bar{\imath}}}
\newcommand{\bj}{{\bar{\jmath}}}

\newcommand{\cfi}{{\phi^*}}
\newcommand{\cg}{{\gamma^*}}
\newcommand{\cT}{T^*}
\newcommand{\cw}{w^*}

\begin{document}

\title{Spin One Matter Fields}

\author{Daniel O.R. Azevedo\thanks{daniel.azevedo@ufv.br}\,, 
Oswaldo M. Del Cima\thanks{oswaldo.delcima@ufv.br}\,,
Thadeu D.S. Dias\thanks{thadeu.dias@ufv.br}\,,\es
Daniel H.T. Franco\thanks{daniel.franco@ufv.br}\,,
Emílio D. Pereira\thanks{emilio.drumond@ufv.br}\, 
and Olivier Piguet\thanks{opiguet@yahoo.com}
\footnote{Permanent address: Pra\c ca Graccho Cardoso, 
76/504, 49015-180 Aracaju, SE, Brazil}
\\[4mm]
\small Universidade Federal de Vi\c cosa (UFV),
Departamento de F\'\i sica\\ 
\small Campus Universit\'ario,
Avenida Peter Henry Rolfs s/n\\
\small 36570-900 Vi\c cosa - MG - Brazil.
}

\date{February 12, 2025}
\maketitle
\begin{abstract}

It is shown how spin one vector matter fields can be coupled 
to a Yang-Mills theory. Such matter fields are defined as 
belonging to a representation $R$ of this Yang-Mills gauge 
algebra $\mathfrak{g}$. It is also required that these fields 
together with the original gauge fields be the gauge fields 
of an embedding total gauge algebra $\mathfrak{g}_{\rm tot}$. 
The existence of a physically consistent Yang-Mills action for 
the total algebra is finally required. These conditions are 
rather restrictive, as shown in some examples: 
non-trivial solutions may 
or may not exist depending on the choice of the original 
algebra $\mathfrak{g}$ and of the representation $R$. 
Some examples are shown, the
case of the initial algebra $\mathfrak{g}$ = 
$\mathfrak{u}(1)\oplus\mathfrak{su}(2)$ being treated in more detail.

\end{abstract}

\newpage

\tableofcontents


\section{Introduction}

Three years after the proposal by Lee and Yang~\cite{Lee-Yang} to
 describe the weak interactions by the exchange of charged  
 spin one ``intermediary bosons'', denoted by $W^\pm$ and 
 a couple of years 
before the creation of the Standard Model~\cite{Cottingham}, 
Arnowitt and Deser~\cite{Arnowitt-Deser} discussed the problem of spin 1 
matter fields, more specifically that of coupling a set of two vector 
fields of electric charge $\pm e$ with the electromagnetic field. 
They showed that the only  consistent way to achieve this is
within an  SU(2) non-Abelian  Yang-Mills theory~\cite{YM}, 
the gauge field multiplet of the latter
being composed by the electromagnetic field together with the 
spin one matter fields.   

A short time before the publication of Ref.\cite{Arnowitt-Deser} 
appeared a quite general proposal by Ogievetsky and 
Polubarinov~\cite{OP1,OP2,OP3,OP4,Ivanov} for the definition of 
higher spin tensorial fields. For the special case of spin 1
vector fields, this amounted to the search of a mechanism for 
eliminating its spin 0 part. For massless vector fields, which is 
our case of interest here, their result is the existence of a 
Yang-Mills theory having these vector fields as its gauge fields. 
This was confirmed in various ways, among them the BRST 
formalism (see~\cite{Barnich} and references to previous works therein).

In the present paper, we propose a definition of 
``spin-one matter fields'' $\f^i_\m$ coupled to a given 
non-Abelian Yang-Mills theory with gauge algebra $\mathfrak{g}$ 
and gauge fields $A^a_\m$.  These fields, which necessarily must 
belong to a larger Yang-Mills theory with gauge algebra 
$\mathfrak{g}_{\rm tot}$ $\supset$ $\mathfrak{g}$,
are required to transform in some 
representation $R$ of the gauge algebra $\mathfrak{g}$. 
We also require that the total algebra $\mathfrak{g}_{\rm tot}$ be minimal, 
in the sense that its dimension is equal to the given number of fields 
$A^a$ and $\f^i$. 
This definition is clearly more restricting than simply demanding the fields 
$\f^i$ to belong, together with the $A^a$, to some larger Yang-Mills theory
with $\mathfrak{g}_{\rm tot}$ $\supset$ $\mathfrak{g}$.
We will look for the whole set of necessary and sufficient 
conditions on the couplings for the existence of a solution
obeying to our requirements in a general situation, 
and apply them to some concrete examples. 
The field content of the present paper is restricted to 
the Yang-Mills fields associated with the initial gauge algebra 
$\mathfrak{g}$ and the spin one matter fields, all taken without mass.

We  begin in Section \ref{charged-fields} by 
reviewing  a simple example, in the BRST framework, 
as a preparation for our further discussion. 
The general scheme is exposed in 
Section \ref{mainsection} and some examples are worked out in
Section \ref{examples}. We conclude with our  final remarks and 
some prospective in the Conclusion Section. 
Some useful facts about Lie Algebra theory are presented in 
Appendix A, as well as a discussion on renormalizability in Appendix B. 

Numerical computations related to Subsections \ref{u1+su2} 
and \ref{s03} were done with 
the help of the software
Mathematica~\cite{Mathematica}. The programs are available as  ancillary 
files accompanying this publication.

\section{The theory of two electrically charged vector fields}
\label{charged-fields}

We give here a short summary of the model proposed 
in~\cite{Arnowitt-Deser},  and as an example in~\cite{OP2}, 
for two charged
vector fields\footnote{Named $W^\pm_\m$ by them, as a reference 
to the weak interaction
intermediate bosons.} $\f_\m^i$, $i=1,2$, interacting 
with the electromagnetic vector potential $A_\m$. Their main result
is that the theory is only consistent if these three 
vector fields are the gauge fields of a Yang-Mills theory 
for the algebra 
$\mathfrak{su}$(2).
We sketch here their construction, but using the BRST 
formalism~\cite{BRS-CMP,BRS-AnnPhys,Tyutin,pig-sor} 
for the discussion of the consistency of the theory.

The vector fields  $\f ^i_\m$ are assumed to be in some 
two-dimensional unitary representation of U(1). 
Then, the (infinitesimal) gauge transformations 
are given by
\eq 
\d_\om A_\m = \pa_\m \om, \quad \d \f_\m^i = \om\,T^i{}_j \f_\m^j,
\eqn{AD-gauge-tr-A}
where $\om(x)$ is the infinitesimal parameter and the matrix $T$ 
is anti-Hermitian\footnote{The special choice $T=i\s_z$ is made 
in~\cite{Arnowitt-Deser}, with charge indexes $i=+,-$.}.
A first condition one has to impose to the theory is to forbid the  
propagation of the longitudinal modes of the vector fields $\f ^i$
in order to preserve the unitarity of the corresponding quantum theory. 
The known way to do it is through 
the imposition of a second gauge invariance, the vector fields transforming under it as
\eq
\d_\eta \f_\m^i = \pa_\m\eta^i + \mbox{possible non-linear terms},
\eqn{AD-gauge-tr-phi}
$\eta^i(x)$ being two new infinitesimal parameters.
The BRST scheme imposes invariance under the BRST transformations $s$ defined as follows:
The vector fields transform as in \equ{AD-gauge-tr-A} and \equ{AD-gauge-tr-phi},
(plus possible non-linear terms)
but with the infinitesimal parameters $\om$ and $\eta^i$ replaced
by anticommuting scalar ghost fields $c$ and $\g^i$, respectively, and the ghosts transforming themselves in such a way to render the BRST operator nilpotent:
\eq 
s^2=0.
\eqn{s-nilpotency}
As it is well known~\cite{BRS-AnnPhys}, this nilpotency defines the 
algebraic structure of the theory. Let us see how this works here.
We first formally put together all the vector fields in a triplet
$(\AA^\a,\,\a=0,1,2)$ =  $(A,\f ^1,\f ^2)$ and similarly for the 
anticommuting ghost fields,
$(\GG^\a,\,\a=0,1,2)$ =  $(c,\g^1,\g^2)$.
We then make the most general 
Ansatz for the  BRST transformations:
\eq
s \AA_\m^\a = \pa_\m \GG^\a + \XX_{\b\g}{}^\a \AA_\m^\b\GG^\g,\quad 
s \GG^\a = -\half \CC_{\b\g}{}^\a \GG^\b\GG^\g.
\eqn{AD-Ansatz-BRST}
We have taken into account the ghost number conservation, the ghost fields  carrying
ghost number 1 and the vector fields and possible other matter fields 
carrying ghost number 0.
It is left as an exercise for the reader to check that the $s$-nilpotency 
condition \equ{s-nilpotency} applied to $\GG$ and then to $\AA$, 
yields the conditions
\eq 
\underset{\a,\b,\g}{\cyclsum}  \CC_{\a\b}{}^\d \CC_{\g\d}{}^\e   = 0,\quad
\XX_{\a\b}{}^\g = \CC_{\a\b}{}^\g.
\eqn{AD-Jacobi}
The first condition on the coefficients $\CC_{\a\b}{}^\g$, together with the 
antisymmetry on their lower indexes -- which stems from  
the anticommutativity of the ghosts $\GG^i$ -- allows us to interpret them as the structure constants of a Lie algebra, with basis elements $\TT_\a$ obeying the commutation rules
\eq 
[\TT_\a,\TT_\b] = \CC_{\a\b}{}^\g \TT_\g.
\eqn{AD-CR}
The second condition \equ{AD-Jacobi} states that the RHS of the first of 
Eqs. \equ{AD-Ansatz-BRST} is the covariant derivative of the ghost $\GG$. 
We can thus already conclude that our vector fields 
are the gauge fields of a Yang-Mills theory~\cite{YM}. 

From these structure constants we can write the Killing form,
\eq 
K_{\a\b} = \CC_{\a\g}{}^\d \CC_{\b\d}{}^\g.
\eqn{AD-Killing}
The terms in $c$ in the BRST transformations of the fields $A$ and $\f$ 
represent the U(1) gauge transformations. We shall require them to be the ones given 
in \equ{AD-gauge-tr-A} (with $\om$ replaced by $c$). This implies the 
supplementary conditions
\eq 
\CC_{0i}{}^0 =  0,\quad 
\CC_{0j}{}^i = -\CC_{j0}{}^i =  T^i{}_j,
\eqn{AD-restrictions}
on the structure constants, where $ T$ is the matrix appearing in
the gauge transformations \equ{AD-gauge-tr-A}.  
The remaining non-vanishing ones can be parametrized as
\eq 
\CC_{ij}{}^0 = \la \e_{ij},\quad \CC_{ij}{}^k = \e_{ij} t^k,
\eqn{AD-CC-diff-0}
where $\e_{ij}$ is the 2-dimensional Levi-Civita tensor 
and $\la$, $t^k$ are three, 
up to now indeterminate, parameters. 
The latter equalities are the consequence of the antisymmetry of 
the structure constants in their lower indexes.

Now, the Jacobi identities \equ{AD-Jacobi} for $(\a,\b,\g,\e)=(0,j,k,n)$
and  for  $(\a,\b,\g,\e)=(0,j,k,0)$ yield, respectively,
\eq
 T^k{}_l\, t^l = 0,\quad \tr T=0.
\eqn{AD-traceT}
One can easily check, using the anti-Hermicity  of $T$, that this implies
the vanishing of $t^i$, hence of $\CC_{ij}{}^k$.
We don't write the remaining Jacobi identities, 
which turn out to hold trivially.
Thus, the consistent BRST transformations finally read
\eq\ba{ll}
s A_\m = \pa_\m c + \la\, \e_{ij}\f_\m^i\g^j, \quad& 
s c = -\dfrac{\la}{2}\e_{ij} \g^i\g^j,\es
s \f ^i_\m 
   =  \pa_\m \g^i + T^i{}_j\lp A_\m\g^j- c\, \f_\m^j \rp ,\quad&
 s \g^i = - T^i{}_j c \g^j,
\ea\eqn{AD-final-s}
with $s^2=0$.
The non-vanishing commutation rules \equ{AD-CR} of the generators of 
the total gauge algebra now explicitly read
\[ 
[\TT_0, \TT_i] = - T^j{}_i \TT_j,\quad [\TT_i, \TT_j] = \la  \e_{ij} \TT_0.
\]
Choosing, without loss of generality, $T=i\s_y$,
one can recognize here
the standard commutations rules of the Lie algebra of 
SU(2):
\eq 
[\TT_{\a-1},\TT_{\b-1}] = \e_{\a\b}{}^\g \TT_{\g-1},\quad \a,\b,\g=1,2,3,
\eqn{CR-SU(2)}
choosing $\la=1$. This is the choice of a positive $\la$ 
which makes the Killing form negative definite.
This can check from the expression
\eq 
\lp K_{\a\b}\rp = -2 \mbox{ diag}\,(1,\la,\la),
\eqn{AD-Killing-su(2)}
of the  Killing form \equ{AD-Killing} before the fixing of $\la$.
Note that, for $\la$ negative, the algebra $\mathfrak{g}_{\rm tot}$
will be that of the non-compact Lie algebra of Sp(2,$\mathbb{R}$).
Thus (for $\la$ positive), we recover the result 
of~\cite{OP2,Arnowitt-Deser}.

\section{Spin one matter in a Yang-Mills theory}\label{mainsection}

\subsection{Gauge invariances and algebraic structure of the 
theory}\label{subsect3.1}

Let us begin with a Yang-Mills gauge theory
associated to a Lie algebra $\mathfrak{g}$, in dimension 
four space-time\footnote{Everything in the present paper, 
except some considerations on renormalization, 
generalizes straightforwardly
to any dimension.}. The algebra may contain an invariant Abelian subalgebra 
- which we shall restrict to be $\mathfrak{u}(1)$, \ie that 
of a U(1) subgroup, in order to avoid too complicated notations:
\eq 
\mathfrak{g} =\mathfrak{u}(1)\oplus \mathfrak{g}_{\rm S},
\eqn{G=U1xGS}
where $\mathfrak{g}_{\rm S}$ is a semisimple Lie 
algebra of dimension $d_{\rm S}$. 
The commutation relations of given basis elements $\tau_0$ and $\tau_a$($a=1,\cdots,d_{\rm S}$) of the Lie algebra $\mathfrak{g}$  read
\eq 
[\tau_0,\tau_a]=0,\quad [\tau_a,\tau_b]=f_{ab}{}^c\tau_c,
\eqn{CR-G}
the $f_{ab}{}^c$'s being the structure constants of 
$\mathfrak{g}_{\rm S}$.
Our aim is to couple, to the gauge fields 
$A^0_\m$ and $A^a_\m$ associated to $\mathfrak{u}(1)$ and 
$\mathfrak{g}_{\rm S}$, a set of vector fields 
("spin 1 matter fields")  multiplets
$\f_\m^{i}$ ($i=1,\cdots,d_R$)
in some representation $R$ of $\mathfrak{g}$, of dimension $d_R$, 
irreducible or not. 
The representation of $\mathfrak{u}(1)$ is characterized 
by a ``hypercharge'' 
$Y_i$ given to each component $\f^i$, and that of 
$\mathfrak{g}_{\rm S}$
by matrices 
$T_a{}^{i}{}_{j}$ obeying the same commutation rules as the generators
$\tau_a$ in \equ{CR-G}. 
Thus, the (infinitesimal) gauge transformations of $A$ and $\f$ 
are given by
\eq\ba{l}
\d A_\m^0 = \pa_\m \om^0 ,\quad 
\d A_\m^a = \pa_\m \om^a + f_{bc}{}^a A_\m^b \om^c, \quad
\d \f_\m^{i} = i Y_i \f_\m^{i} \om^0 
- T_a{}^{i}{}_{j} \f_\m^{j} \om^a,
\ea\eqn{g-trans-A-c}
where $\om^0$ and $\om^a$ are the infinitesimal parameters  associated to
$\mathfrak{u}(1)$ and $\mathfrak{g}_{\rm S}$, respectively

We shall express the  invariance under the gauge 
transformations 
\equ{g-trans-A-c} as an invariance under nilpotent BRST 
transformations, denoted here by  $\scc$, which,
for $A^0$ and $A^a$, are just the transformations 
\equ{g-trans-A-c}
with the infinitesimal parameters replaced by the anticommuting 
Faddeev-Popov fields $c^0$, $c^a$:
\eq\ba{lll}
\scc A_\m^0 = \pa_\m c^0 ,&\quad 
\scc A_\m^a = \pa_\m c^a + f_{bc}{}^a A_\m^b c^c, &\quad
\scc \f_\m^{i} = i Y_i \f_\m^{i} c^0 
- T_a{}^{i}{}_{j} \f_\m^{j} c^a,\es
\scc c^0 =0, &\quad \scc c^a = -\frac12 f_{bc}{}^a c^b c^c &\es
\scc^2=0,&&
\ea\eqn{BRST-A-c}
The algebraic structure of the theory is expressed through the
nilpotency of the BRST operator $\scc$~\cite{BRS-AnnPhys}. 
Indeed, it is easy to check that applying twice $\scc$ to $A$ and $c$ 
yields  the correct
Jacobi identities for the structure constants and commutation
relations of the representation matrices $T$:
\eq
\underset{a,b,c}{\cyclsum} f_{ab}{}^d f_{cd}{}^e   = 0,\quad\quad\quad
[T_0,T_a]=0,\quad [T_a,T_b]=f_{ab}{}^c\, T_c,
\eqn{Jacobi-f}
where $\cyclsum$ is the cyclic sum symbol.

It is well known (see~\cite{OP1,OP2,OP3,OP4}
as well as~\cite{Barnich} and other references in 
that paper) that any consistent theory involving vector fields must be a 
Yang-Mills theory.
In particular, the vector fields $\f$ will transform as
\[
\d\f_\m^{i}=\pa_\m \eta^{i} + \mbox{non-linear terms},
\]
under new gauge transformations of infinitesimal parameter $\eta^{i}$,
in order to be free of non-physical longitudinal modes.

This means that we must consider an extension of $\mathfrak{g}$,
the ``total algebra'' $\mathfrak{g}_{\rm tot}$, of dimension  
determined by the 
total number of gauge fields $A^0$, $A^a$ and $\f^i$: 
\eq 
d_{\rm tot} = 1+d_S+d_R.
\eqn{3.x} 
Let us introduce a basis of $\mathfrak{g}_{\rm tot}$, 
$\TT_\a$ ($\a=1,\cdots,d_{\rm tot}$), obeying commutation rules
\eq
[\TT_\a,\TT_\b]=\CC_{\a\b}{}^\g\TT_\g,
\eqn{CR-Gtot}
the structure constant $\CC_{\a\b}{}^\g$ being antisymmetric 
in their lower indexes and obeying the Jacobi identities
\eq 
\underset{\a,\b,\g}{\cyclsum} \CC_{\a\b}{}^\d \CC_{\g\d}{}^\e   = 0.
\eqn{Jacobi-CC}
Denoting by $w_{i}$ the basis elements of
$\mathfrak{g}_{\rm tot}$ associated to the gauge fields $\f^{i}$, 
we make the identification
\eq 
\{\TT_\a,\,\a=1,\cdots,d_{\rm tot}\}= 
\{\tau_0,\, \tau_a,\, w_{i};\ a=1,\dots,d_{\rm S},\,i=1,\cdots,d_R\},
\eqn{multilabels1}
The same index convention will be followed for any tensorial quantity with indexes $\a$, $\b$, \etc.

The BRST transformations corresponding to the algebraic structure defined by
\equ{CR-Gtot} and \equ{Jacobi-CC} read
\eq\ba{l}
s \AA_\m^\a = \pa_\m \GG^\a + \CC_{\b\g}{}^\a \AA_\m^\b \GG^\g,\quad
s\GG^\a = -\half\, \CC_{\b\g}{}^\a\, \GG^\b\GG^\g,\quad s^2=0,
\ea\eqn{BRST-AA-GG}
where the gauge fields $\AA$ and ghost fields $\GG$ are given in components by
\eq 
\{\AA^\a\} = \{A^0,A^a,\f^{i}\},\quad
\{\GG^\a\} = \{c^0,c^a,\g^{i}\},
\eqn{def-AA-GG}
the $\g^{i}$'s being the ghosts associated to the fields $\f^{i}$.

Now, the choice of the total gauge algebra 
$\mathfrak{g}_{\rm tot}$ will be submitted to
the condition that the original gauge transformations parametrized by 
$\om^0$, $\om^a$ will be maintained as they are shown in \equ{g-trans-A-c}
or, equivalently, \equ{BRST-A-c}.
These requirements, together with the condition \equ{3.x}, represent our definition of the $\f^{i}$'s being
``spin 1 matter fields  in a representation of the gauge 
algebra $\mathfrak{g}$''. 
Thus $\mathfrak{g}_{\rm tot}$ is not any Lie algebra 
extension of  $\mathfrak{g}$. Our condition indeed implies the vanishing 
or the fixing of part of its structure constants\footnote{In this Section we shall systematically do 
an abuse of notation writing, \eg $\CC_{ab}{}^{i}$ instead of 
$\CC_{1+a,1+b}{}^{d_S+1+i}$  \label{footnotation}}:
\eq\ba{l} 
\CC_{0a}{}^0 =  \CC_{ab}{}^0 =  
\CC_{0a}{}^b = \CC_{0a}{}^{i} = \CC_{ab}{}^{i} =  0,
\quad \CC_{ab}{}^{c} =f_{ab}{}^{c},\es
\CC_{0i}{}^0 = \CC_{0i}{}^a = \CC_{ai}{}^0 =  \CC_{ai}{}^b =
 0,\quad
 \CC_{0i}{}^{j} = -iY_i\d_{i}^{j},\quad 
 \CC_{ai}{}^{j} = T_a{}^{j}{}_{i}.
\ea\eqn{CC=0}
We shall refer at \equ{CC=0} as the 
``spin one particle representation'' (S1PR) condition.
It is the generalization of the conditions 
\equ{AD-restrictions} of
the charged fields  model. 
We note that the equations in the first line also mean 
that the initial algebra \equ{G=U1xGS} 
must be a subalgebra of the total one.

Introducing the notations
\eq 
\CC_{ij}{}^0 = S_{ij}{}^0,\quad
\CC_{ij}{}^a = S_{ij}{}^a,\quad
\CC_{ij}{}^{k} = t_{ij}{}^{k},
\eqn{notation_S-T}
we may write the detailed BRST transformations as
\eq\ba{l}
s A_\m^0 = \pa_\m c^0 + S_{ij}{}^0 \f_\m^{i}\g^{j},\quad
s A_\m^a = \pa_\m c^a +f_{bc}{}^a A_\m^b c^c
+ S_{ij}{}^a \f_\m^{i}\g^{j},\es
s\f_\m^{i} = \pa_\m \g^{i} 
  + T_a{}^{i}{}_{j}(A_\m^a\g^{j}-\f_\m^{j}c^a)
 - i Y_i  (A_\m^0\g^{i}-\f^{i}_\m c^0)
 +t_{j k}{}^{i} \f^{j}_\m\g^{k},\es
 s c^0 = -\frac12 S_{ij}{}^0 \g^{i}\g^{j},\hspace{14mm} 
s c^a =-\frac12 f_{bc}{}^a c^b c^c -\frac12 S_{ij}{}^a \g^{i}\g^{j}
\es
s\g^{i} = -T_a{}^{i}{}_{j}c^a \g^{j}
+ i Y_i c^0 \g^{i}
-\frac12 t_{j k}{}^{i} \g^{j}\g^{k}, 
\ea\eqn{detailed-BRST}
as well as the commutation relations of the generators $\tau_0$, 
$\tau_a$ and $w_{i}$ of the total algebra $\mathfrak{g}_{\rm tot}$:
\eq\ba{l}
[\tau_0,\tau_a] = 0,\quad [\tau_0,w_{i}] =-i Y_i w_{i},\es
[\tau_a,\tau_b] = f_{ab}{}^c \tau_c,\quad 
[\tau_a,w_{i}] = T_a{}^{j}{}_{i} w_{j},\es
[w_{i},w_{j}] =  S_{ij}{}^0 \tau_0 + S_{ij}{}^a \tau_a
+  t_{i j}{}^{k} w_{k}.
\ea\eqn{detailed_CR}
From the Jacobi identities \equ{Jacobi-CC} we find the following 
identities that the non-zero structure constants must satisfy. 
(We list only the non-trivially satisfied ones).

For $(\a,\b,\g,\e)$ = $(0,a,i,k)$, $(0,i,j,0)$, $(0,i,j,a)$,
$(i,j,0, l)$ we find, respectively,
\eq\ba{ll}
T_a{}^{k}{}_{i}(Y_k-Y_i)=0,&\quad S_{ij}{}^0(Y_i+Y_j)=0,\es
S_{ij}{}^a(Y_i+Y_j)=0,&\quad t_{ij}{}^{l}(Y_i+Y_j-Y_l)=0.
\ea\eqn{Y-conserv}
These represent four hypercharge conservation laws: $T_a{}^{k}{}_{i}$, 
$S_{ij}{}^0$, $S_{ij}{}^a$ and  $t_{ij}{}^{l}$ 
are different from zero only if $Y_k=Y_i$, $Y_i+Y_j=0$,  
$Y_i+Y_j=0$ and $Y_l=Y_i+Y_j$, respectively.
 An important consequence of the first of Eqs. \equ{Y-conserv} is that 
the representation $R_{\rm S}$ of $\mathfrak{g}_{\rm S}$, 
defined by the matrices $T_a$, factorizes 
as a direct sum of representations, each one being characterized 
by a particular value of the hypercharge. 
Decomposing further each of these representations in irreducible ones, 
one gets the full decomposition
$R_{\rm S}$ = $\oplus_{n=1}^{n_{\rm max}} R_n$, each  irreducible representation
$R_n$  being characterized by a hypercharge $Y_n$ and a set of 
dimension $d_n$ matrices $\Tup[n]_a{}^{i_n}{}_{j_n}$: 
\eq 
T_a = \mbox{Block-diagonal}
\lp \Tup[1]_a,\Tup[2]_a,\cdots,\Tup[n_{\rm max}]_a\rp
\eqn{T-reducibility}

For $(\a,\b,\g,\e)$ = $(a,b,c,e)$ and  $(a,b,i,j)$, we recover 
 the Jacobi identity and commutation relations \equ{Jacobi-f}.

The Jacobi identity \equ{Jacobi-CC} for 
$(\a,\b,\g,\e)$ = $(a,i,j,0)$, $(a,i,j,b)$, $(a,i,j,l)$, 
yield, respectively,
\eq\ba{l}
S_{kj}{}^0 T_a{}^{k}{}_{i} 
+ S_{ik}{}^0 T_a{}^{k}{}_{j} =0,\es
S_{ij}{}^d f_{da}{}^b 
+ S_{kj}{}^b T_a{}^{k}{}_{i} 
+ S_{ik}{}^b T_a{}^{k}{}_{j} =0,\es
t_{kj}{}^{l} T_a{}^{k}{}_{i} 
+ t_{ik}{}^{l} T_a{}^{k}{}_{j}
-t_{ij}{}^{k} T_a{}^{l}{}_{k}  =0.
\ea\eqn{inv-tensors}
These three identities express the invariance of the tensors 
$S_{ij}{}^0$, $S_{ij}{}^b$ and $t_{ij}{}^{l}$ under 
the action of the algebra $\mathfrak{g}_{\rm S}$, 
more precisely under the action 
of the representation matrices
$T_a{}^i{}_j$ and $({\rm ad}_a)_b{}^c=f_{ba}{}^c$.

Finally, for  $(\a,\b,\g,\e)$ = $(i,j,k,0)$, $(i,j,k,a)$, 
$(i,j,k,l)$, we find the three identities
\eq\ba{l}
\underset{i,j,k}{\cyclsum} t_{ij}{}^{l}S_{kl}{}^0 =0,\qquad
\underset{i,j,k}{\cyclsum} t_{ij}{}^{l}S_{kl}{}^a =0,\es
\underset{i,j,k}{\cyclsum}\lp t_{ij}{}^{n}t_{kn}{}^{l}
-  i Y_l S_{ij}{}^0 \d^{l}_{k} 
+S_{ij}{}^a T_a{}^{l}{}_{k} \rp = 0.
\ea\eqn{Jacobi-like}

Let us summarize here the conditions we have imposed in order
to achieve a consistent coupling of the spin one matter fields:
\begin{enumerate}
\item The original $\mathfrak{g}$ gauge fields together with 
the spin one vector fields are the gauge fields of a 
Yang-Mills theory with  gauge algebra 
$\mathfrak{g}_{\rm tot}$ containing  $\mathfrak{g}$ as a subalgebra.
\item The spin one vector fields  belong to a representation of the
initial gauge algebra $\mathfrak{g}$, as implemented by the S1PR conditions
\equ{CC=0}.
\item  There exists a physically relevant Yang-Mills action 
(see Subsection \ref{invar-action}),
invariant under the gauge transformations defined by
 $\mathfrak{g}_{\rm tot}$.
 By this we mean in particular that the action must involve 
the spin one matter fields in a non-trivial way, and also that
its kinetic term must obey the positivity conditions necessary for
the unitarity of the corresponding quantum theory. 
\end{enumerate}
 As we will see in some of the examples discussed below, 
this set of  conditions is rather restrictive.

\subsection{Invariant action}\label{invar-action}

An important quantity we will need is the invariant Killing form of the total 
gauge algebra $\mathfrak{g}_{\rm tot}$:
\eq 
K_{\a\b} = \CC_{\a\g}{}^\d \CC_{\b\d}{}^\g,\quad K_{\b\a}=K_{\a\b}.
\eqn{K-form}
Its components are 
\eq\ba{l}
\ba{ll}
K_{00} = -\sum_{i=1}^{d_R} Y_i^2 ,\quad&
K_{0a} = - i\sum_{i=1}^{d_R} Y_i T_a{}^i{}_i,\quad
K_{ab} = f_{ac}{}^d f_{bd}{}^c + \tr(T_a T_b),\es 
K_{0i} = - i\sum_{j=1}^{d_R} Y_j t_{ij}{}^j,\quad&
K_{ai} = T_a{}^k{}_j t_{ik}{}^j,
\ea\\[7mm]\hspace{2mm}
K_{ij} = - (S^aT_a)_{ij} - (S^aT_a)_{ji} + i(Y_j-Y_i)S_{ij}{}^0
+ t_{il}{}^k t_{jk}{}^l. 
\ea\eqn{Gtot-Killing}
We see that the Killing form may be  
non-degenerate\footnote{\Ie the full algebra 
$\mathfrak{g}_{\rm tot}$  is semisimple, as stated by Cartan's
criterion for semisimplicity~\cite{Ramond}.}
only if at least one of the three
tensors $S^0$, $S^a$ or $t$ is present. 

In the non-degenerate case, 
the Yang-Mills gauge invariant action constructed with the fields $A$ 
and $\f$ reads
\eq 
S_{\rm YM} = \NN\int d^4 x\, K_{\a\b}\FF^\a_{\m\n}\FF^{\b\,\m\n},
\eqn{YM-action}
where $\NN$ is a normalization factor and $\FF^\a_{\m\n} =
\pa_\m\AA_\n^\a-\pa_\n\AA_\m^\a 
+ \CC_{\b\g}{}^\a \AA_\m^\b\AA_\n^\g$ is the curvature 
(field strength) of the gauge field \equ{def-AA-GG}. In components:
\eq\ba{l}
\FF_{\m\n}^0 = \pa_\m A_\n^0-\pa_\n A_\m^0 + S_{ij}{}^0 \f^i_\m\f^j_\n,  \es
\FF_{\m\n}^a = \pa_\m A_\n^a-\pa_\n A_\m^a
+ f_{bc}{}^a A_\m^b A_\n^c + S_{ij}{}^a\f_\m^i\f_
\n^j,\es 
\FF_{\m\n}^i = \pa_\m \f_\n^i-\pa_\n \f_\m^i 
+ T_a{}^i{}_j (A_\m^a \f_\n^j-A_\n^a \f_\m^j)  
-iY_i  (A_\m^0 \f_\n^i-A_\n^0 \f_\m^i)
+t_{jk}{}^i \f_\m^j \f_\n^k .
\ea\eqn{curv-fullalg}
The curvature transforms covariantly under the BRST transformations:
\eq 
s \FF^\a = \CC_{\b\g}{}^\a \FF^\b\GG^\g,
\eqn{BRST-FF}
so that the action \equ{YM-action} is obviously invariant due to the Killing form being 
an invariant tensor in the adjoint representation.

Clearly, the third condition given at the end of the latter Subsection requires the positive (or negative) definiteness of the Killing metric.

\subsubsection*{Degenerate case:}\label{degenerate-case}

If the Killing form is degenerate, the construction of an action 
may be jeopardized. Such a situation occurs if the
three tensors $S^0$, $S^a$ and $t$ vanish: the Killing form components
$K_{0i}$, $K_{ai}$ and $K_{ij}$ indeed vanish
in this case. It then follows from the 
commutation relations \equ{detailed_CR} that 
the generators $\tau_0$, $w_i$ and $w^*_\bi$ span an 
ideal $\mathfrak{h}$, 
\ie an invariant subalgebra, of the  
total algebra $\mathfrak{g}_{\rm tot}$. Moreover one easily 
checks that this ideal is the maximal
solvable ideal (in other words, the radical) of $\mathfrak{g}_{\rm tot}$ 
and that the latter decomposes itself into
the semidirect sum of the semi-simple algebra $\mathfrak{g_{\rm S}}$ 
(see \equ{G=U1xGS})
generated by the $\tau_a$'s, and the radical
$\mathfrak{h}$\footnote{See Appendix \ref{topicsLie} 
for definitions.}:
\eq 
\mathfrak{g}_{\rm tot} = \mathfrak{su}(2)\ltimes\mathfrak{h}.
\eqn{semidirect-gtot}
The invariant action  \equ{Action-I-1/2} with the 
- now degenerate - Killing metric
will depend only on the original gauge fields $A^0$ and $A^a$ 
(see \equ{curv-fullalg}). 
It would be desirable 
to complete it by terms involving the fields $\f^i$.
Such a completion should involve the Yang-Mills field
$\FF^i_{\m\n}$ given in \equ{curv-fullalg}, the BRST transform of the latter
being  $s\FF^{i} =  T_a{}^{i}{}_{j}(F^a\g^{j}-\FF^{j}c^a)
 - i Y_i  (F^0\g^{i}-\FF^{i} c^0)$.  Obviously, 
there is no way to build an invariant with it.

\subsection{Remark on the renormalization of the theory}

The construction presented in Subsection \ref{subsect3.1} 
has been done at the classical level. The result is a Yang-Mills gauge theory
associated to a certain Lie algebra $\mathfrak{g}_{\rm tot}$.
A general result~\cite{tHooft,tHooft-Veltman,BRS-CMP,BRS-AnnPhys} 
is that such a Yang-Mills theory -- in four dimensions --
is renormalizable and thus exists perturbatively  as a quantum theory whose 
$\hbar=0$ limit is the given classical theory.  
(The Standard Model of particle interactions is the most well-known example.)
For this, some conveniently chosen  gauge fixing condition has 
to be implemented. We discuss in Appendix \ref{app-ren} the special choice
of the Landau gauge, for which some non-renormalization theorems 
hold\cite{antighost,pig-sor}, which in our case assure the explicit 
preservation to  all orders of the 
S1PR conditions \equ{CC=0}. But this does not mean of course
that any other choice for the gauge fixing should be excluded, 
the only difference being in the possible renormalization of 
these conditions, without changing the physical content of the theory.

\section{Examples}\label{examples}

\subsection{Initial Lie algebra is $\mathfrak{so}(N)$}
In this example, 
we take the vector fields $A^a$ ($a=1, \dots, N(N-1)/2$)
as the gauge fields of the Lie algebra $\mathfrak{so}(N)$, 
and the  vector fields $\f^i$ 
($i=1,\cdots,N$) transforming in the adjoint representation
of $\mathfrak{so}(N)$. 

A basis for this Lie algebra $\mathfrak{so}$(N) may be chosen as 
a set of $N(N-1)/2$ 
antisymmetric $N\times N$ matrices $\tau_{[mn]}$, indexed by the  $N(N-1)/2$ 
antisymmetric pairs of indexes $[mn]$, with their matrix elements given by
\eq 
\tau_{[mn]}{}^{ij} = \d^i_m\d^j_n- \d^j_m\d^i_n,
\eqn{so(n)-matrices}
and obeying the commutation rules 
\eq 
[\tau_{[mn]},\tau_{[pq]}] = \d_{mq}\tau_{[np]} + \d_{np}\tau_{[mq]} 
- \d_{mp}\tau_{[nq]}  - \d_{nq}\tau_{[mp]}.
\eqn{so(n)-CR}
It will be convenient to substitute the antisymmetric pairs $[mn]$, etc.,
by indexes $a$ = $a([m,n])$ defined recursively by:
\[\ba{ll}  
a([1,2]) = 1,&\es 
a([m,n]) = a([m,n-1]) + 1,\quad&(1\leq m\leq N-1,\ m+2\leq n\leq N),\es
a([m,m+1]) = a([m-1,N]) + 1,\quad&(2\leq m\leq N-1).
\ea\]
We can thus write the BRST transformations as in 
\equ{detailed-BRST}.

\subsubsection*{Case $S = -T$}

We shall make here the following restrictions on the 
tensorial coefficients:
\eq 
T^a{}_{ij}=\tau^a_{ij},\quad
S^a{}_{ij} = -\tau^a{}_{ij},\quad t_{ijk}=0.
\eqn{S=T&t=0}
The first condition just states that the spin one matter field is 
chosen in the adjoint representation.
For the second condition, we used the facts that, in the present case, 
we don't need to worry on the indexes' positions, 
and that the $\tau_a{}^i{}_j$'s and $S^a{}_{ij}$s 
are both antisymmetric in $i,j$. With this, \equ{detailed-BRST} simplifies to
\eqa
&s A_\m^a = \pa_\m c^a + f_{bc}{}^a A_\m^b c^c 
- \tau^a{}_{ij}\f_\m^i\g^j,\quad
&s c^a = -\half\lp f_{bc}{}^a c^b c^c 
- \tau^a{}_{ij}\g^i\g^j \rp,\nonumber\\
&s\f_\m^i = - c^a \tau_a{}^i{}_j \f_\m^j 
+ \pa_\m \g^i + A_\m^a \tau_a{}^i{}_j\f^j
,\quad
&s\g^i = - c^a \tau_a{}^i{}_j\g^j .
\eqan{BRST-so(n)}
The commutation rules 
\equ{CR-Gtot} and the Killing form \equ{K-form},
\equ{Gtot-Killing} for the full Lie algebra now read
\eq 
[\tau_a,\tau_b] = f_{ab}{}^c \tau_c,\quad
[\tau_a,w_i] = \tau_a{}^j{}_i w_j,\quad
[w_i,w_j] = -\tau^a{}_{ij}  \tau_a ,
\eqn{CR-Gtot-so(n)}

\eq
K_{ab} = k_{ab} + g_{ab} ,\quad 
K_{ai}=K_{ia} = 0, \quad K_{ij} = 2C_{ij},
\eqn{K-form-so(n)}
where $k_{ab}=f_{ac}{}^d f_{bd}{}^c=-2(N-2)\d_{ab}$ is 
the Killing form of $\mathfrak{so}(N)$, 
$g_{ab}=\tr(\tau_a \tau_b)=-2\d_{ab}$ is its trace form and 
$C_{ij}=(\tau^a \tau_a)_{ij} = -(N-1) \d_{ij}$ is its 
 quadratic Casimir operator
in the adjoint representation~\cite{Haber}.
Hence 
\eq 
K_{\a\b} = -2(N-1) \d_{\a\b},
\eqn{K-form-so(n+1+1}
which is the Killing form of $\mathfrak{so}(N+1)$.
Note that the signal's choice in the second of Eqs. \equ{S=T&t=0} insures the
negative definiteness of the Killing form, hence the compactness of the total Lie
algebra $\mathfrak{g}_{\rm tot}$ and thus the existence of a physically relevant action.

It turns out that  $\mathfrak{g}_{\rm tot}$ is in fact 
$\mathfrak{so}$(N+1). Indeed, 
the Lie algebra basis elements $\tau_a$ and $w_i$ together with their
commutation relations \equ{CR-Gtot-so(n)} may be represented by the 
$(N+1)\times(N+1)$ antisymmetric matrices $T_{[\a,\b]}$ 
($\a,\b=1\cdots,N+1$), defined by
\eq\ba{l}
T_{[m,n]}{}^{ij} = \tau_{[m,n]}{}^{ij},\quad
T_{[m,n]}{}^{i,N+1} = 0,\quad\quad
T_{[m,N+1]}{}^{ij} = 0,\quad
T_{[m,N+1]}{}^{i,N+1} = \d^i_m,\es
(m,n,i,j=1,\cdots,N),
\ea\eqn{matrices T_a}
where we have come  back to the notation in terms of 
antisymmetric pairs of indexes as in \equ{so(n)-matrices}.
It can easily be checked that these matrices form a basis for the 
adjoint representation of $\mathfrak{g}_{\rm tot}$ = 
$\mathfrak{so}(N+1)$.

\subsection{Initial Lie algebra is 
$\mathfrak{u}(1)\oplus\mathfrak{su}(2)$}\label{u1+su2}

In this example\footnote{The  
Mathematica notebook containing the calculations is available as the ancillary file ``Example G=U1xSU2.nb''accompanying the present arXiv submission.} 
spin-one matter is described by a vector field multiplet 
$\f^i_\m$ ($i=1,\cdots,d_I$) 
in some irreducible representation of isospin $I$ of dimension $d_I=2I+1$, of 
$\mathfrak{su}(2)$, with hypercharge $Y$, and its complex 
conjugate $\cfi{}^\bi_\m$  ($\bi=1,\cdots,d_I$)
in the conjugate representation, with hypercharge $-Y$. 
The BRST transformations \equ{detailed-BRST} and 
the commutation relations\equ{detailed_CR} now read
\eq\ba{l}
s A_\m^0 = \pa_\m c^0 +
 S_{i\bj}{}^0 (\f_\m^{i}\cg^{\bj}- \cfi_\m^{\bj}\g^{i}),\quad
s A_\m^a = \pa_\m c^a +\e_{bc}{}^a A_\m^b c^c
+ S_{i\bj}{}^a (\f_\m^{i}\cg^{\bj}- \cfi_\m^{\bj}\g^{i}),\es
s\f_\m^{i} = \pa_\m \g^{i} 
  + T_a{}^{i}{}_{j}(A_\m^a\g^{j}-\f_\m^{j}c^a)
 - i Y  (A_\m^0\g^{i}-\f^{i}_\m c^0),\es
s\cfi_\m^{\bi} = \pa_\m \cg^{\bi} 
  + \cT_a{}^{\bi}{}_{\bj}(A_\m^a\cg^{\bj}-\cfi_\m^{\bj}c^a)
 + i Y  (A_\m^0\cg^{\bi}-\cfi^{\bi}_\m c^0),\es
 s c^0 = -S_{i\bj}{}^0 \g^{i}\cg^{\bj},\hspace{14mm} 
s c^a = - \frac12 \e_{bc}{}^a c^b c^c - S_{i\bj}{}^a \g^{i}\cg^{\bj}
\es
s\g^{i} = -T_a{}^{i}{}_{j}c^a \g^{j} + i Y c^0 \g^{i}, 
\es
s\cg^{\bi} = -\cT_a{}^{\bi}{}_{\bj}c^a \cg^{\bj}- i Y c^0 \cg^{\bi}, 
\ea\eqn{BRST-Y1/2}
and

\eq\ba{lll}
[\tau_0,\tau_a] = 0,\quad &
[\tau_0,w_{i}] =-i Y_i w_{i},\quad& [\tau_0,\cw_{\bi}] =i Y \cw_{\bi},\es
[\tau_a,\tau_b] = \e_{ab}{}^c \tau_c,\quad& 
[\tau_a,w_{i}] = T_a{}^{j}{}_{i} w_{j},\quad&
[\tau_a,\cw_{i}] = \cT_a{}^{\bj}{}_{\bi} \cw_{\bj},\es
[w_{i},w_{j}] =0,\quad&  [\cw_{\bi},\cw_{\bj}] =0,\quad&
[w_{i},\cw_{\bj}] =  S_{i\bj}{}^0 \tau_0 + S_{i\bj}{}^a \tau_a.
\ea\eqn{CR-Y1/2}
$\e_{ab}{}^c$ is the completely antisymmetric
 Levi-Civita tensor, with $\e_{12}{}^3=1$.
Note that the tensors  $t_{ij}{}^k$, $t_{i\bj}{}^k$, $t_{\bi\bj}{}^k$, 
as well as $S_{ij}{}^{(0,a)}$ and $S_{\bi\,\bj}{}^{(0,a)}$, 
are zero  
due to the Jacobi identities \equ{Y-conserv} expressing 
hypercharge conservation. In the same way, the representation 
matrices are block-diagonal according to \equ{T-reducibility}, 
the two blocks being  
made of anti-Hermitian 
$d_I\times d_I$ - matrices  $T_a$ ($a=1,2,3$)
and their complex conjugates $T^*_a$, respectively. The matrices 
$T_a$ and $T^*_a$ obey  the same commutation relations as the 
fundamental  $\tau_a$ in \equ{CR-Y1/2}.

\subsubsection{Isospin 1/2 and 1}

Let us begin with the case of $R$ being the fundamental representation, 
\ie that of 
isospin 1/2. The basis representation matrices are proportional to 
the Pauli matrices:
\eq 
T_a = -\frac{i}{2}\s_a,\quad \cT_a = \frac{i}{2}\s^*_a, \quad a=1,2,3.
\eqn{T_a-Y1/2}  
We look now for solutions to the conditions
listed in \equ{inv-tensors}
and \equ{Jacobi-like} for the tensors $S_{i\bj}{}^0$ and $S_{i\bj}{}^a$.
The first two conditions in \equ{inv-tensors} already determine
them up to two arbitrary constants $\ka$ and $\la$:\footnote{The 
higher or lower positions of the indexes $a,\,i,\,\bi$, \etc are a 
purely aesthetic choice.}
\eq 
S_{i\bj}{}^0 =-i \ka \,\d_{i\bj},\quad 
S_{i\bj}{}^a = \la\, \cT{}^a{}_{i\bj} = - \la\, T^a{}_{\bj i}.
\eqn{S0-Sa-mu-la}
Finally, the third condition in \equ{Jacobi-like} fixes $\la$ 
in terms of the hypercharge and $\ka$, which remain as the 
two free parameters of the model:
\eq 
\la=-\frac{4}{3}Y\, \ka.
\eqn{lambda}
The Killing form matrix \equ{K-form}-\equ{Gtot-Killing} reads then 
\eq 
\lp K_{\a\b}\rp = - \lp\ba{cccccccc}
4 Y^2&0 &0 &0 &0 &0 &0 &0 \\
0 &3 &0 &0 &0 &0 &0 &0 \\
0 &0 &3 &0 &0 &0 &0 &0 \\
0 &0 &0 &3 &0 &0 &0 &0 \\
0 &0 &0 &0 &0 &0 &4 Y\ka &0 \\
0 &0 &0 &0 &0 &0 &0 &4 Y\ka \\
0 &0 &0 &0 &4 Y\ka &0 &0 &0 \\
0 &0 &0 &0 &0 &4 Y\ka &0 &0 
\ea\rp
\eqn{Killing-I=1/2}
The tensor components of the total algebra $\mathfrak{g}_{\rm tot}$ 
are labeled by the index set 
\[
\Lac\a=1,\cdots,d_{\rm tot}=4+2d_I\Rac = 
\Lac\{1\},\{a=1,\cdots,3\},\{i=1,\cdots,d_I\},\{\bi=1,\cdots,d_I\}\Rac.
\]
An abuse of notation analog to the one indicated in the footnote
\ref{footnotation} of Section \ref{mainsection} will be done here
as well.

The curvature components may be read from \equ{curv-fullalg}:
\eq\ba{l}
\FF_{\m\n}^0 = \pa_\m A_\n^0-\pa_\n A_\m^0 
+ S_{i\bj}{}^0 \f^i_\m\cfi^\bj_\n,  \es
\FF_{\m\n}^a = \pa_\m A_\n^a-\pa_\n A_\m^a
+ f_{bc}{}^a A_\m^b A_\n^c 
+ S_{i\bj}{}^a\f_\m^i\cfi_\n^\bj,\es 
\FF_{\m\n}^i = \pa_\m \f_\n^i-\pa_\n \f_\m^i 
+ T_a{}^i{}_j (A_\m^a \f_\n^j-A_\n^a \f_\m^j)  
-iY  (A_\m^0 \f_\n^i-A_\n^0 \f_\m^i) \es
\FF_{\m\n}^{*\bi} = \pa_\m \cfi_\n^\bi-\pa_\n \cfi_\m^\bi 
+ T^*_a{}^\bi{}_\bj (A_\m^a \cfi_\n^\bj-A_\n^a \cfi_\m^\bj)  
+iY  (A_\m^0 \cfi_\n^\bi-A_\n^0 \cfi_\m^\bi) ,
\ea\eqn{curv-I=1/2}
and the Yang-Mills part of the action from \equ{YM-action}:
\eq 
S_{\rm YM} = -\NN\int d^4 x\,\lp
4Y^2\FF_{\m\n}^0 \FF^{\m\n}_0
+3\FF_{\m\n}^a \FF^{\m\n}_a
+8Y\ka\,\FF_{\m\n}^{*\bi} \FF^{\m\n}_i\rp.
\eqn{Action-I-1/2}
This result in particular shows that the Killing form is in 
fact negative definite, provided $Y$ and $\ka$ are chosen with 
the same sign,
and the procedure thus yields a physically consistent action, 
as required by the conditions spelled out at the end of 
Subsection \ref{subsect3.1}.

We ask now the question on the nature of the total Lie algebra 
$\mathfrak{g}_{\rm tot}$
generated by the basis elements $\tau^0$, $\tau^a$, $w_i$ and 
$w^*_\bi$ obeying the commutation rules 
\equ{CR-Y1/2}. From the non-degeneration and negative definiteness of
the Killing form we deduce that it is semisimple and compact. 
Hence it must be the direct sum of simple, compact Lie 
algebras.
Since its dimension is equal to 8, and since the simple compact 
Lie algebras of dimension up to 8 are $\mathfrak{su}(2)$, 
$\mathfrak{su}(3)$ and $\mathfrak{so}(4)$, of dimensions 
3, 8 and 6, respectively, we conclude that the total Lie 
algebra is $\mathfrak{su}(3)$.
This can also be checked explicitly from the commutation relations
\equ{CR-Y1/2} and the expressions \equ{T_a-Y1/2} and 
\equ{S0-Sa-mu-la}  for the tensors $T_a$, $S^0$ and $S^a$. 
One easily sees indeed that the basis elements 
$\{\tau_0,\tau_a,w_i,w^*_\bi\}$ of the total algebra
can be expanded in terms of the SU(3) Gell-Mann 
matrices $\la_\a$~\cite{Ramond}:
\eq\ba{ll}
\tau_0 = -i\dfrac{Y}{\sqrt{3}}\la_8,\quad&
\tau_a = -i\dfrac12 \la_a\ \ (a=1,2,3),\es
w_1 = -i\sqrt{\dfrac{Y\ka}{6}}(\la_4+i\la_5),\quad&
w_2 = -i\sqrt{\dfrac{Y\ka}{6}}(\la_6+i\la_7),\es
w^*_1 = -i\sqrt{\dfrac{Y\ka}{6}}(\la_4-i\la_5)\quad&
w^*_2 = -i\sqrt{\dfrac{Y\ka}{6}}(\la_6-i\la_7).
\ea\eqn{Cartan-Weyl}
Note that, up to multiplicative factors, this is just the 
Cartan-Weyl basis of $\mathfrak{su}(3)$.
The field redefinitions  corresponding to this 
change of basis are given by
\eq\ba{lll}
\hat\AA{}^a = A^a,\quad a=1,2,3,&&\es
\hat\AA{}^4 = \g(\f^1+\cfi^1),\quad& \hat\AA{}^5 = i\g(\f^1-\cfi^1),&\es
\hat\AA{}^6 = \g(\f^2+\cfi^2),\quad& \hat\AA{}^7 = i\g(\f^2-\cfi^2),
\quad& \hat\AA{}^8 = \dfrac{2Y}{\sqrt{3}} A^0,
\ea\eqn{troca-campos}
where $\g=\sqrt{2Y\ka/3}$. One checks that, in the new basis
$\la_\a$, the action \equ{Action-I-1/2} now reads
\eq
S_{\rm YM} 
= -\NN'\int d^4 x\,\hat{K}_{\a\b}\hat\FF{}^\a\hat\FF{}^\b,
\eqn{New-action}
where $\hat K_{\a\b}$ = $12\,\d_{\a\b}$ is the SU(3) Killing form in
the basis  $\la_\a$, and the new field strengths 
$\hat\FF{}^\a$ are related to the former ones 
in the same way as the $\hat\AA$ fields are related to the 
$A$ and $\f$ fields in \equ{troca-campos}.

In the case where the matter vector field representation is the 
direct sum of the isospin 1 and its conjugate, we find  
very similar results. Without giving every detail, we just mention 
that the matrices $T_a$ are now of the $I=1$ three-dimensional representation,
Eqs. \equ{S0-Sa-mu-la} still holds, but \equ{lambda} is replaced by
$\la=-Y\ka$.
In what concerns the total Lie algebra, its dimension here is 
equal to 10, and a similar argument shows that the total 
Lie algebra is $\mathfrak{so}(5)$.

\subsubsection{Higher isospins}\label{higher-isospin}

A similar result could be expected to hold for isospin 
values greater than 1. But it turns out, at least for the  isospin values in
the interval $3/2\leq I \leq10$ which we have explored, that the only 
solutions are those with $Y\ka=\la=0$, where $\ka$ and $\la$ are
the coefficients shown in \equ{S0-Sa-mu-la}. 
The Killing metric vanishes,
we are in the degenerate situation depicted at the end of Section 
\ref{mainsection}. There are two cases, the first one with $Y\not=0$, $\ka=0$,
and the second one with $Y=0$, $\ka$ vanishing or not. In both cases the 
generators $\tau^0$, $w_i$ and $w^*_\bi$ span a radical $\mathfrak{h}$ of 
$\mathfrak{g}_{\rm tot}$, with the decomposition \equ{semidirect-gtot},
and there is no invariant action involving all fields 
$A^0$, $A^a$ and $\f^i$.

Of course, a negative result such as those mentioned here does 
not exclude the existence of another representation of the same dimension, 
not necessarily irreducible, such that a nontrivial algebra 
$\mathfrak{g}_{\rm tot}$ does exist. For instance, in the case 
of $I=9/2$, the construction has yielded a
trivial solution. But it is well known~\cite{Georgi-Glashow} that 
the algebra $\mathfrak{su}(5)$ contains 
$\mathfrak{u}(1)\oplus\mathfrak{su}(2)$ as a subalgebra, the remaining 
20 generators belonging to a representation 
of dimension 20 of the latter.

\subsection{Initial Lie algebra is $\mathfrak{so}(3)$
}\label{s03}

A more simple example is shown here in a summarized way\footnote{The  
interested reader may wish to consult the Mathematica notebook 
containing the detailed calculations, available 
as the  ancillary file ``Example so3 to su(3).nb''
accompanying the present arXiv submission.}. 
The starting algebra $\mathfrak{g}$ is
$\mathfrak{so}(3)$, with structure constants given as 
the Levi-Civita tensor: $f_{ab}{}^c = \epsilon_{abc}$.
The spin one field $\f^a_\m$ is chosen to
transform in the representation of isospin $I=2$. A basis 
for the generators of $\mathfrak{so}(3)$ in this representation 
is given by the matrices $T_a$, $a=1,2,3$:
\eq
\lp\begin{array}{ccccc}
 0&0&0&1&0 \\ 0&0&-1&0&0 \\ 0&1&0&0&0 \\-1&0&0&0&\sqrt{3}\\ 
  0&0&0&-\sqrt{3}&0 
 \ea\rp,\quad
  \lp\ba{ccccc}
0&0&1&0&0 \\ 0&0&0&1&0 \\ -1&0&0&0&-\sqrt{3} \\ 0&-1&0&0&0 \\
 0&0&\sqrt{3}&0&0
\ea\rp,\quad
\lp\ba{ccccc}
0&-2&0&0&0 \\ 2&0&0&0&0 \\ 0&0&0&-1&0 \\ 0&0&1&0&0 \\ 0&0&0&0&0
\end{array}\right).
\eqn{so3-generators}
Solving the Jacobi identities \equ{Jacobi-CC} for the structure constants 
$\CC_{\a\b}{}^\g$ ($\a,\b,\g=1,\cdots,8$) of the total algebra, 
taking into account the S1PR conditions \equ{CC=0}, 
yields a set of families of  structure constants 
(antisymmetric in the first two indixces), indexed by a parameter $\la$:
\eq\ba{lllll}
\CC_{34}{}^5 = 2
,&\quad \CC_{35}{}^4 = -2
,&\quad \CC_{36}{}^7 = 1
,&\quad \CC_{37}{}^6 = -1
,&\quad \CC_{45}{}^3 = -2\la/\sqrt{3},\es
\CC_{46}{}^2 = \la/\sqrt{3}
,&\quad \CC_{47}{}^1 = \la/\sqrt{3}
,&\quad \CC_{56}{}^1 = -\la/\sqrt{3}
,&\quad \CC_{57}{}^2 = \la/\sqrt{3}
,&\quad \CC_{67}{}^3 = -\la/\sqrt{3},\es
\CC_{68}{}^2 = -\la
,&\quad \CC_{78}{}^1 = \la,&&&
\ea\eqn{non0-st-const}
where only the independent and non-vanishing one are displayed. The corresponding 
Killing form reads
\eq 
\lp K_{\a\b}\rp = \mbox{diag}(-12,\ -12,\ -12,\ 4\sqrt{3}\la,
\ 4\sqrt{3}\la,\ 
4\sqrt{3}\la,\ 4\sqrt{3}\la,\ 4\sqrt{3}\la).
\eqn{Killing-s03tosu4}
Imposing to the free parameter the condition $\la<0$ makes this form 
negative definite.
With this choice the total algebra is compact and semi-simple, 
hence equal to
$\mathfrak{su}(3)$, the only such algebra of dimension 8. 
The solutions with 
$\la>0$ yield a nondefinite Killing form and must then be 
discarded for physical reasons (see the third of conditions 
displayed at the end of Section \ref{subsect3.1}).

\subsection{An example where $\mathfrak{g}_{\rm tot}$ is not 
simple}\label{notsimple}

The matter field multiplet $\f_\m$ is chosen to transform under the initial Lie 
algebra\footnote{The relevant equations are those of Section 
\ref{mainsection}, all terms containing $Y$ or a subscript 
or superscript 0 being excluded.} $\mathfrak{g}$ 
in the adjoint representation, given by the matrices
$T_a{}^i{}_j=f_{a j}{}^i$, 
where the $f_{ab}{}^c$'s are the structure constants 
of $\mathfrak{g}$. The indexes $a,b, i,j,$ \etc are now all 
algebra indexes, varying between 1 and $d_g$, the dimension of $\mathfrak{g}$.
The tensors $S_{ij}^a$ and $t_{ij}^k$ appearing in the BRST transformations
\equ{detailed-BRST} and the commutation relations 
\equ{detailed_CR} 
being invariant tensors now in the adjoint representation, 
are proportional to the structure constants: $S_{ij}^a$ = $Sf_{ij}^a$, 
$t_{ij}^k$ = $tf_{ij}^k$. The total algebra's Jacobi identities \equ{inv-tensors} and 
\equ{Jacobi-like} obviously follow from those satisfied by the $f_{ab}{}^c$.
The commutation relations \equ{detailed_CR} thus read
\eq
[\tau_a,\tau_b] = f_{ab}{}^c \tau_c,\quad 
[\tau_a,w_{b}] = f_{ab}{}^c w_{c},\quad
[w_{a},w_{b}] =  Sf_{ab}{}^c \tau_c +  tf_{ab}{}^{c} w_{c}.
\eqn{CR-ad}
Redefining the generators $w_a$ as $w'_a=w_a-\tau_a$ leads to the 
commutation rules
\eq
[\tau_a,\tau_b] = f_{ab}{}^c \tau_c,\quad 
[\tau_a,w'_{b}] = 0,\quad
[w'_{a},w'_{b}] =  (S+t-1)f_{ab}{}^c \tau_c +  tf_{ab}{}^{c} w'_{c}.
\eqn{CR-ad-new}
Choosing the parameters $S$, $t$ such that $S+t=1$ leads to a
situation where the total algebra  $\mathfrak{g}_{\rm tot}$ is not simple,
but the direct sum of $\mathfrak{g}$ by itself:
\eq 
\mathfrak{g}_{\rm tot} =  \mathfrak{g}\oplus \mathfrak{g}.
\eqn{semisimple}

\section{Conclusion}

Given a gauge theory based on some given gauge algebra
 $\mathfrak{g}$,
we have shown how to implement the conditions we have proposed in order to define spin-one matter fields as belonging to 
a given representation of this gauge algebra. 
This led to the construction of an extension $\mathfrak{g}_{\rm tot}$ of the initial gauge algebra, submitted to a set of conditions summarized at the end of Subsection \ref{subsect3.1}, and to look for a physically relevant corresponding action.

We then applied these conditions 
to three examples. The first one is that of the 
initial gauge algebra $\mathfrak{g}$ being $\mathfrak{so}(N)$. 
We have restricted ourselves to the 
adjoint representation for the matter field $\f^i_\m$, 
and made a special Ansatz
for the choice of the structure constants 
(see the last two of Eqs. \equ{S=T&t=0}). The result is the 
total algebra being $\mathfrak{so}(N+1)$ -- a rather obvious embedding 
of the initial algebra. 

The second  example, where the initial algebra is $\mathfrak{u}(1)\oplus\mathfrak{su}(2)$,  is far less trivial and  potentially more 
interesting for physics. The representations of  $\mathfrak{su}(2)$
we have taken into account are
the direct sums of a (generally complex) irreducible representation 
of isospin $I$  and its conjugate, carrying $\mathfrak{u}(1)$
hypercharges $Y$ and $-Y$, respectively. Of all the isospin values 
$I=1/2,\,1,\,\cdots,\, 10$ considered in the numerical 
evaluations, only
the values $I=1/2$ and $1$ have yielded  non-degenerate, compact 
total algebras $\mathfrak{g}_{\rm tot}$, which turned out to be $\mathfrak{su}(3)$ and
$\mathfrak{so}(5)$, respectively.  
For the other values of $I$, the total algebra is not semisimple, 
more precisely, it is the semidirect sum of the initial 
algebra $\mathfrak{su}(2)$ by a non-Abelian radical 
$\mathfrak{h}$, the basis of which is given by the 
$\mathfrak{u}(1)$ generator $\tau_0$ and the generators 
$w_i$ and $\cw_i$ associated 
to the matter fields $\f_\m^i$ and their conjugates. 
As these latter solutions do not provide an action  
involving the spin-one matter fields, they must be discarded.
The physically relevant solutions are those which produce a 
simple or semi-simple algebra.

The third example starts from the Lie algebra 
$\mathfrak{so}(3)$, the spin one matter field being in the isospin 2 
representation. The construction yields $\mathfrak{su}(3)$ as 
the total algebra.

We can conclude, in particular from the second example, that 
our conditions are very
selective. Solutions with a simple or semisimple total algebra 
may be present or not, depending on the choice of the initial algebra 
 and its representation $R$ for the matter multiplet 
$\f^i_\m$. Moreover nothing guarantees that the resulting 
total algebra is compact, hence having a 
definite Killing (positive or negative) form that allows 
the construction of an action 
compatible with unitarity, \ie such that the matrix
of the propagator residues at each physical pole is positive definite. 

We restricted ourselves to the original Yang-Mills fields 
$A^a_\m$ and the matter fields $\f^i_\m$, all taken as massless. 
An important generalization, not considered in the present paper
but deserving further investigation, 
would be the introduction of spin-one-half
matter fields, as well as of suitable spin-zero Higgs fields 
in order to produce mass for some of (or all) the matter and 
Yang-Mills fields.

We have to recall that the construction shown in this paper has used 
from the start the result originally presented in the papers 
\cite{OP1,OP2,OP3,OP4} and in subsequent references which may be found 
in \cite{Barnich}, namely  the necessity to consider a Yang-Mills type
gauge theory in order to describe consistently the propagation of spin one particles. It is in this framework that we look for a gauge theory 
where the ``spin-one matter fields'' are required to belong to a specific,
given
representation of the original gauge group. As we have seen, a 
solution may or may not exist.

A final remark is that this construction is a realization of 
the ``embedding'' of a Lie algebra into a larger Lie algebra, as
defined by 
Gell-Mann, Ramond and Slansky (GRS)~\cite{Gell-Mann,Slansky} in the 
context of the Grand Unification  Theories (see also~\cite{Ramond}). 
Beyond being explicit and constructive,
our approach can be thought of as a reverse of GRS's one: instead 
of looking for the possible embedded subalgebras $\mathfrak{g}$
of a given simple Lie algebra $\mathfrak{g}_{\rm tot}$, we have
shown how to find $\mathfrak{g}_{\rm tot}$, given a Lie algebra 
$\mathfrak{g}$ and given a representation of the latter, 
with the help of the 
conditions we have proposed for the definition of a spin one particle.
Note that the total Lie algebra $\mathfrak{g}_{\rm tot}$ may not always be 
simple, as shown in Subsection \ref{notsimple}, a situation not 
contemplated in Refs.~\cite{Gell-Mann,Slansky,Ramond}.

\noindent{\bf Acknowledgments.} 
We thank very much Prof. Evgeny Ivanov for informing us about the 
Refs.~\cite{OP1,OP2,OP3,OP4,Ivanov}.
This study was financed in part by the Coordenação de Aperfeiçoamento de Pessoal de Nível Superior – Brasil (CAPES) – Finance Code 001.
O.P. thanks his colleagues from the UFV physics department
for their always friendly welcome.

\appendix
\section*{Appendices}

\section{Topics on Lie algebras}\label{topicsLie}

General information may be found in the book~\cite{Ramond}. The more specific properties reported in this Appendix are found 
in~\cite{Ostrowski} and~\cite{Onishchik}.

\subsubsection*{Semidirect sums of Lie algebras:}
Let us consider two Lie algebras $\mathfrak{g}$ and $\mathfrak{h}$ with no common element other than zero, and an homomorphism 
$\s:\mathfrak{g} \to \mbox{Der}\,\mathfrak{h}$, where 
$\mbox{Der}\,\mathfrak{h}$ is the set of  
{\it derivative operators} acting on  $\mathfrak{h}$.
The latter means that such an operator, $\s$, acts as
\[\ba{l}
\s(g_1 g_2)=\s(g_1) g_2+g_1\s(g_2),\quad \quad \forall 
g_1,g_2 \in \mathfrak{g}, \es
\s(g)[h_1,h_2] = [\s(g) h_1,h_2]+[h_1,\s(g) h_2],\quad \forall 
g\in \mathfrak{g} \mbox{ and }  h_1,h_2 \in \mathfrak{h}.
\ea\]
We denote by $(g,h)$ the pair of one element of $\mathfrak{g}$ and 
one element of $\mathfrak{h}$. 
Then, the {\it semidirect sum} of $\mathfrak{g}$ by $\mathfrak{h}$,
denoted  $\mathfrak{g}\ltimes\mathfrak{h}$, is 
defined~\cite{Ostrowski}  by the bracket
\eq
[(g_1,h_1),(g_2,h_2)] = \lp[g_1,g_2]\,,\,[h_1,h_2]
+\s(g_1)h_2 - \s(g_2)h_1\rp.
\eqn{def-semidirect}
Identifying the derivative operator $\s(g)$ as  
the adjoint  action of $\mathfrak{g}$ on $\mathfrak{h}$,
\[
\s(g)h = [g,h],
\]
yields the algebra \equ{detailed_CR} in the case of vanishing tensors
$S_{ij}{}^0$ and $S_{ij}{}^a$, if  $\mathfrak{g}$ and $\mathfrak{h}$ 
are identified with the algebras spanned by the basis
$\{\tau_0,\tau_a\}$ and $\{w_i\}$, respectively.

\subsubsection*{Solvable algebra:}
A Lie algebra $\mathfrak{g}$ is called solvable if the sequence of commutators and nested commutators
\[
[\mathfrak{g},\mathfrak{g}]\supseteq 
[[\mathfrak{g},\mathfrak{g}],[\mathfrak{g},\mathfrak{g}]],\supseteq 
[[[\mathfrak{g},\mathfrak{g}],[\mathfrak{g},\mathfrak{g}]],[[\mathfrak{g},\mathfrak{g}],[\mathfrak{g},\mathfrak{g}]]]\supseteq  \cdots,
\]
where $[\mathfrak{g},\mathfrak{g}]$ denotes the commutator of any element
of $\mathfrak{g}$ with any other element of $\mathfrak{g}$,
is finite and terminates with the zero subalgebra.

\subsubsection*{Levi's decomposition:}
A theorem proved by Levi~\cite{Levi,Onishchik} states that any Lie 
algebra $\mathfrak{g}$ on the real or complex numbers can be decomposed 
as the semidirect product of a semisimple $\mathfrak{g}_{\rm S}$ by 
the  maximal solvable ideal of $\mathfrak{g}$, its radical rad($\mathfrak{g}$):
\eq 
\mathfrak{g} = \mathfrak{g}_{\rm S} \ltimes {\rm rad}(\mathfrak{g}).
\eqn{Levi-theorem}


\section{Renormalization in the Landau gauge}
\label{app-ren}

In general, conditions obeyed by the parameters of a theory,
such as the ``S1PR'' conditions \equ{CC=0},  are expected to be 
renormalized by quantum corrections. We show here that the S1PR 
conditions are actually explicitly preserved to all orders of renormalized perturbation theory if one works in the Landau gauge. In order to do so, we take advantage of the antighost 
equation~\cite{antighost,pig-sor} which holds in this gauge.

Let us first recall the functional formalism used in the 
BRS algebraic renormalization program~\cite{pig-sor}.
The implementation of the Landau gauge is achieved by adding to the 
gauge invariant action \equ{YM-action} a trivially 
BRST-invariant gauge fixing action
\eq
S_{\rm gf} = s\int d^4 x\bar{\GG}_\a \partial^\mu \AA_\mu^\a
 = \int d^4 x \lp\mathcal{B}_\a \partial^\mu \AA_\mu^\a 
- \bar{\GG}_\a \pa^\m(\partial_\mu \GG^\a + \CC_{\g\b}{}^\a \AA_\mu^\g\GG^\b)
 \rp,
\eqn{gauge-fixing}
where one has 
introduced a new set of fields, namely the antighosts 
$\bar{\GG}_\a$ and the Lagrange multipliers $\BB_\a$, which 
transform as BRST doublets:
\eq 
s\bar{\GG}_\a=\BB_\a,\quad  s\BB_\a=0.
\eqn{BRS-GG-BB}
One recovers the Landau gauge condition through the functional identity
\eq 
\frac{\d S}{\d \mathcal{B}_\a} = \partial^\mu \AA_\mu^\a,
\eqn{gauge-cond}
where
\eq 
S = S_{\rm YM} + S_{\rm gf} +  S_{\rm ext},
\eqn{total-action}
is the total action. The last term,
\eq 
S_{\rm ext} = \int d^4x \lp \rho^\mu_\a s\AA_\mu^\a 
+ \sigma_\a s\GG^\a\rp,
\eqn{S_ext}
defines the coupling of  external sources $\rho$ and $\s$ 
to the  $\AA$ and $\GG$ BRST transforms 
\equ{BRST-AA-GG} or \equ{detailed-BRST}, 
introduced in order to control their renormalization.

The BRST invariance of the theory is expressed functionally,
at the classical level, by the Slavnov-Taylor identity
\eq
\mathcal{S}(S) = \int d^4x\left( 
\frac{\d S}{\d \rho_\a^\mu}\frac{\d S}{\d \AA^{\a}_\mu} + 
\frac{\d S}{\d \sigma_\a}\frac{\d S}{\d \GG^\a} + 
\mathcal{B}_\a \frac{\d S}{\d \bar{\GG}_\a} \right) = 0.
\eqn{st-id} 
Another functional identity, the ghost equation
\eq 
\frac{\d S}{\d \bar{\GG}_\a} + 
\partial_\m \frac{\d S}{\d \rho_{\m\a}} = 0,
\eqn{ghost-eq}
is a consequence of \equ{gauge-cond} and \equ{st-id}.

The restriction to the Landau gauge 
as defined by \equ{gauge-cond}, instead of a general 
linear covariant gauge,  gives us an additional functional
identity: the antighost equation 
\eq 
\int d^4 x \left(\frac{\d S}{\d \GG^\g} 
- \CC_{\b\g}{}^\a \bar{\GG}_\a \frac{\d S}{\d \mathcal{B}_\b} \right) 
= \D^{cl}_\g
\eqn{antighost-eq}
where
\eq 
\D^{cl}_\g = \int d^4 x\, \CC_{\b\g}^\a \lp -\rho_\a^\m \AA_\m^\b + \sigma_\a \GG^\b \rp
\eqn{classical-breaking}
Renormalizability amounts to showing that there is a solution of 
the Slavnov identity \equ{st-id} together with the gauge condition 
\equ{gauge-cond}, with the action $S$ substituted by the
generating functional $\Gamma[\AA,\GG,\BB,\bar{\GG},\rho,\s]$ 
of the amputated 
1-particle irreducible Green functions. The solution is obtained 
as a (formal) power series in $\hbar$ (equivalently: in the number of 
loops in the Feynman graphs). At zeroth order, $\Gamma=S$.
Renormalizability was proved in the general case of a Yang-Mills 
theory with scalar and spinor fields, in the absence of the
Adler–Bardeen non-Abelian gauge 
anomaly~\cite{tHooft,tHooft-Veltman,BRS-AnnPhys,pig-sor}.

The antighost equation \equ{antighost-eq}, with $S$ substituted 
by $\Gamma$, was proven to hold at all orders, 
too~\cite{antighost,pig-sor}. 
It is a peculiarity of the Landau gauge choice, and its usefulness 
resides in its right-hand side being linear in the dynamical 
fields $\AA$ and $\GG$, which implies the non-renormalization 
of its coefficients, the structure constants $ \CC_{\g\b}^\a$. 
As a consequence, the S1PR conditions \equ{CC=0} we have 
imposed on these coefficients remain stable in the 
renormalization process. As we have just said, this 
non-renormalization property is peculiar to the Landau gauge: 
for any different gauge choice the structure constants 
and their relations may be renormalized, but the physical 
content of the theory remains unchanged.



\end{document}